\documentclass[twocolumn,aps,pra]{revtex4-1} 

\usepackage{amsmath,amssymb,amsfonts,bm}
\usepackage{graphicx,epstopdf}  
\usepackage{color}
\usepackage{extarrows}
\usepackage{enumitem}
\usepackage{empheq}

\renewcommand{\d}{{\rm d}}

\newcommand{\w}{\omega}
\newcommand{\wti}{\widetilde}

\newcommand{\B}{\mbox{\tiny B}}
\newcommand{\s}{\mbox{\tiny S}}

\newcommand{\tS}{\mbox{\tiny S}}
\newcommand{\T}{\mbox{\tiny T}}

\newcommand{\dg}{\dagger}
\newcommand{\la}{\langle}
\newcommand{\ra}{\rangle}

\renewcommand{\bm}{\pmb}

\newcommand{\be}{\begin{equation}}
\newcommand{\ee}{\end{equation}}
\newcommand{\bsube}{\begin{subequations}}
\newcommand{\esube}{\end{subequations}}
\newcommand{\Eq}[1]{Eq.\,(\ref{#1})}
\newcommand{\Eqs}[1]{Eqs.\,(\ref{#1})}
\newcommand{\Fig}[1]{Fig.\,\ref{#1}}

\newcommand{\RN}[1]{%
  \textup{\uppercase\expandafter{\romannumeral#1}}%
}

\begin{document}

\title{\color{black}{A statistical quasi-particles thermofield theory with Gaussian environments: System--bath entanglement theorem for nonequilibrium correlation functions}
}

\author{Yao Wang} \email{wy2010@ustc.edu.cn}
\author{Zi-Hao Chen}%
\author{Rui-Xue Xu}
\author{Xiao Zheng}
\author{YiJing Yan} \email{yanyj@ustc.edu.cn}
\affiliation{
Department of Chemical Physics,
 University of Science and Technology of China, Hefei, Anhui 230026, China
}

\date{\today}
\begin{abstract}
For open quantum systems,  Gaussian environmental dissipative effect can be represented by statistical quasi-particles, namely dissipatons. 
 We exploit this fact to establish the dissipaton thermofield theory. 
The resulting generalized Langevin dynamics of absorptive and emissive thermofield operators are effectively noise--resolved.
{\color{black} The system--bath entanglement theorem} is then readily followed between a important class of nonequilibrium steady--state correlation functions.
All these relations are validated numerically. A simple corollary is the transport current expression, which exactly recovers the result obtained from the nonequilibrium Green's function formalism.

\end{abstract}
\maketitle

\section{Introduction} Thermal effects are of vital importance in various realms of
physics, ranging from elementary particles \cite{Wei743357,Gro8143} to  quantum devices \cite{Bro16045005,Ber21025003}  to biological molecules \cite{Pan20259}.
For open systems, thermal effects arise from the coupling environments and dictate the system--environment correlations, the
thermodynamics \cite{Kir35300,Gon20154111,Gon20214115}
and transport properties \cite{Mei922512,Hau08,Gru1624514,Ber21025003}
in quantum impurities.
 Understanding the thermal effects in strongly correlated systems is closely related to the coherent  manipulations in spintronic and superconducting quantum interference devices \cite{Gar04,Cle101155,Soa14825,Har201184,Koc81380}.
There are two major theoretical approaches
to open quantum systems.
One is the nonequilibrium Green's function (NEGF) formalism, which is a type of correlation function approach, related to various quantum transport properties \cite{Sch61407,Kel651018}.
Another is the real--time quantum dissipation method, such as  the semi-group quantum master equations \cite{Lin76119,Gor76821,Ali87},
Feynman--Vernon influence functional path integral theory \cite{Fey63118}, and its derivative equivalence the hierarchical equations
of motion formalism \cite{Tan906676,Tan06082001,Yan04216,
Xu05041103,Xu07031107,Jin08234703}.
These traditional quantum dissipation theories primarily focus on reduced system dynamics, without revealing relations between various correlation functions, in particular those involving thermal environment modes.

In this work, we develop an alternative method, the dissipaton thermofield (DTF) theory.
This is a real--time dynamic method, which also readily reveals relations between different correlation functions.
The DTF theory goes with a generalized nonequilibrium Langevin equation description of the thermofield modes.
We exploit the statistical quasi--particles (named as dissipatons) to characterize the non-Markovian influence of environments.
%
%
The proposed DTF theory extends the conventional thermofield method \cite{Sch61407,Kel651018,Ume95} with the quasi-particle dynamics picture.
The resulting generalized Langevin dynamics of absorptive and emissive thermofield operators are effectively noise--resolved.
This greatly facilitates establishing  the relations between various nonequilibrium Green's functions or correlation functions, as demonstrated in this work.
This is named as system--bath entanglement theorem for nonequilibrium steady--state correlation functions.
Here, the phrase of ``system--bath entanglement'' justifies the context where there exists coherence between  system and bath, constituting a type of many--particle composite \cite{Du20034102}.
All above mentioned relations are validated numerically.

Moreover, the DTF dynamics agrees  well with the dissipaton equation of motion (DEOM) theory \cite{Yan14054105,Yan16110306,Wan20041102}, a second--quantization version of the hierarchical equations of motion formalism.
It is also noticed that DEOM is practically equivalent to the emerging
discrete pseudo-mode semi-group form of the reduced core--system dynamics method
\cite{Tam18030402,Tam19090402,Che19123035,Lam193721}.
Therefore, the proposed DTF theory that integrated with DEOM will serve as a versatile tool for the thermal effects in strongly correlated systems.
For brevity, we set throughout this paper $\hbar=1$ and $\beta_{\alpha}\equiv 1/(k_BT_{\alpha})$, with $T_{\alpha}$ being the temperature of the $\alpha$th reservoir and $k_B$ the Boltzmann constant.

\section{Background}
Let us start with the total system--plus--reservoirs composite Hamiltonian,
\be\label{HSB_boson}
 H_{\T}=H_{\tS}+H_{\tS\B}+h_{\B},
\ \ \text{with} \ \
 H_{\tS\B}=\sum_{\alpha,u}\hat Q_{u}\hat F_{\alpha u} .
\ee
Both the system Hamiltonian $H_{\tS}$
 and the dissipative system modes $\{\hat Q_u\}$ are arbitrary,
whereas the hybrid reservoir bath modes $\{\hat F_{\alpha u}\}$
assume to be linear.
This together with noninteracting reservoir  model of $h_{\B}=\sum_{\alpha}h_{\alpha}$
constitute the Gaussian environment ansatz \cite{Wei12,Kle09}.
The environmental influence is fully characterized by
the interacting bath reservoir correlation functions ($t\geq 0$):
\be\label{Fcorr_boson}
\begin{split}
{\bm c}_{\alpha}(t) = \{c_{\alpha uv}(t)\equiv \la\hat F^{\B}_{\alpha u}(t)\hat F^{\B}_{\alpha v}(0)\ra_{\B}\} .
\end{split}
\ee
Here, $\hat F^{\B}_{\alpha u}(t)\equiv
e^{ih_{\B}t}\hat F_{\alpha u}e^{-ih_{\B}t}
=e^{ih_{\alpha}t}\hat F_{\alpha u}e^{-ih_{\alpha}t}
$
and $\la(\,\cdot\,)\ra_{\B}\equiv{\rm tr}_{\B}[(\,\cdot\,)
\rho^{0}_{\B}]$ with $\rho^0_{\B}=\otimes_{\alpha} [e^{-\beta_{\alpha}h_{\alpha}}/{\rm tr}_{\B}(e^{-\beta_{\alpha}h_{\alpha}})]$.
 The corresponding response functions are
\be\label{phit}
 {\bm \phi}_{\alpha}(t)=\big\{\phi_{\alpha uv}(t)
\equiv i\la[\hat F^{\B}_{\alpha u}(t),\hat F^{\B}_{\alpha v}(0)]\ra_{\B}
\big\},
\ee
satisfying $\phi_{\alpha uv}(t)=i [c_{\alpha uv}(t)-c^{\ast}_{\alpha uv}(t)]$.
Inversely, ${\bm c}_{\alpha}(t)$
can also determined by ${\bm \phi}_{\alpha}(t)$  via
the fluctuation--dissipation theorem \cite{Wei12,Kle09,Zhe121129,Yan05187}.

Consider now the $H_{\T}$--based Heisenberg picture
of the hybrid bath modes. It is easy to obtain \cite{Du20034102, Du212155}
\be\label{QLE_boson}
   \hat F_{\alpha u}(t)=\hat F^{\B}_{\alpha u}(t)
 -\sum_v \int^{t}_{0}\!\d\tau\,\phi_{\alpha uv}(t-\tau)\hat Q_v(\tau),
\ee
with the Langevin random force $\hat F^{\B}_{\alpha u}(t)$
on the local system dissipative mode $\hat Q_{u}$;
see \Eq{HSB_boson}.
In other words, \Eq{QLE_boson} is the precursor to conventional quantum Langevin equation.
%
It together with $[\hat F^{\B}_{\alpha u}(t), \hat Q_{v}(0)]=0$ give rise to the system--bath entanglement
theorem for response functions, which is a type of input--output relations in the total composite space \cite{Du20034102}.
On the other hand, $\la\hat F^{\B}_{\alpha u}(t) \hat Q_{v}(0)\ra\neq 0$. 
To obtain the nonequilibrium steady--state correlation function type input--output relations for such as
\bsube
\begin{align}
\label{Cinput_boson}
 {\bm C}_{\tS\tS}(t)=&\{C^{\tS\tS}_{uv}(t)\equiv \la \hat Q_{u}(t)\hat Q_v(0)\ra\},
 \\ \label{CalpS}
 {\bm C}_{\alpha\tS}(t)
= &\{C^{\alpha\tS}_{uv}(t)
 \equiv\la \hat F_{\alpha u}(t)\hat Q_{v}(0)\ra\},
\end{align}
\esube
\Eq{QLE_boson} is insufficient since it cannot distinguish the absorptive and emissive contributions.
More specifically,  from \Eq{QLE_boson} we have 
\be\label{CSalp_boson_final}
 {\bm C}_{\alpha \tS}(t)
={\bm X}_{\alpha\tS}(t) -\int_{0}^{t}\!{\rm d}\tau
\,{\bm \phi}_{\alpha}(t-\tau) \cdot{\bm C}_{\tS\tS}(\tau),
\ee
with
\be
{\bm X}_{\alpha\tS}(t)=\{
X^{\alpha\tS}_{uv}(t)\equiv \la\hat F^{\B}_{\alpha u}(t)\hat Q_{v}(0)\ra\},
\ee
to be further resolved. 
To that end, we  exploit the statistical quasi-particles picture, which is used in the DEOM theory \cite{Yan14054105}, and obtain
\be\label{termI1}
 {\bm X}_{\alpha\tS}(t)
 =\sum_k {\bm X}_{\alpha\tS k}(t),
\ee
with
\be \label{g2}
X_{uvk}^{\alpha \tS}(t)\!\equiv\! \la  \hat f^{\B}_{\alpha u k}(t)\hat Q_v(0)\ra\!=\!\la  \hat f_{\alpha u k}(0)\hat Q_v(0)\ra e^{-\gamma_{\alpha k}t}.\!
\ee
Here, $\{\gamma_{\alpha k}\}$ originates from the exponential decomposition of the interacting bath correlations reading \cite{Yan16110306}
 \be \label{FFdecomp}
c_{\alpha u v}(t)=\sum_{k}\eta_{\alpha uv k}e^{-\gamma_{\alpha k}t}.
\ee
Evidently, to establish the aforementioned correlation function type input--output relations, the key step is to formulate $X_{uvk}^{\alpha \tS}(0)$ in terms of ${\bm C}_{\tS\tS}(t)$ [cf.\,\Eq{Cinput_boson}] and ${\bm c}_{\alpha}(t)$ [cf.\,\Eq{Fcorr_boson}].
We address this issue within the scope of DTF theory to be elaborated as follows.
\begin{figure}
\includegraphics[width=1\columnwidth]{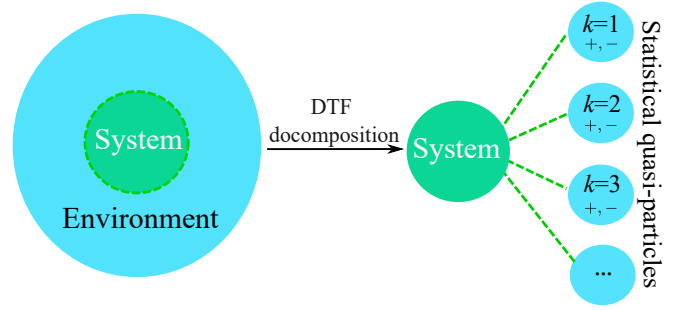}
  \caption{An illustrative depiction of the DTF decomposition, exemplified with the case of single reservoir and single coupling mode. Before the decomposition,
the hybrid reservoir mode obeys \Eq{QLE_boson}. After the decomposition, statistical quasi--particles evolve as \Eq{QLE2}.
}
\label{fig1}
\end{figure}

\section{Dissipaton thermofield theory}
\subsection{Ansatzes}
The proposed DTF theory is based on
the dissipaton decomposition of the hybrid reservoir modes,
as schematically represented in \Fig{fig1}. 
 There are three basic ingredients:

\noindent(\emph{i})
 \emph{Dissipaton decomposition ansatz:} The hybrid reservoir modes can be decomposed into dissipatons as
 \bsube\label{eq9}
\be\label{F_f_boson}
 \hat F_{\alpha u}=\sum_{k}  \hat f_{\alpha uk}
\ee
with [cf.\,\Eqs{Fcorr_boson} and  (\ref{FFdecomp})]
\be\label{ff_boson}
\begin{split}
 \la\hat f_{\alpha u k}^{\B}(t)\hat f_{\alpha' v k'}^{\B}(0)\ra_{\B}
=\delta_{\alpha \alpha'}\delta_{kk'}\eta_{\alpha u v k}e^{-\gamma_{\alpha k}t},
\\
 \la\hat f_{\alpha' v k'}^{\B}(0)\hat f_{\alpha u k}^{\B}(t)\ra_{\B}=\delta_{\alpha \alpha'}\delta_{kk'}\eta^{\ast}_{\alpha uv \bar k}e^{-\gamma_{\alpha k}t},
\end{split}
\ee
\esube
satisfying  $\la\hat f_{\alpha' v \bar k}^{\B}(0)\hat f_{\alpha u \bar k}^{\B}(t)\ra_{\B}=\la\hat f_{\alpha u k}^{\B}(t)\hat f_{\alpha' v k}^{\B}(0)\ra_{\B}^{*}$, the time--reversal relation, where $\gamma^{\ast}_{\alpha k}\equiv \gamma_{\alpha \bar k}$
%
and $\hat f_{\alpha u k}^{\B}(t)\equiv e^{ih_{\B}t}
\hat f_{\alpha u k}e^{-ih_{\B}t}$. This defines dissipatons that are  statistically independent
diffusive environmental modes, exploited previously in the DEOM theory  \cite{Yan14054105}.

\noindent(\emph{ii})
 \emph{Thermofield dissipatons ansatz}: Each $\hat f_{\alpha u k}$ consists of
an absorptive ($+$) and an emissive ($-$) parts,
\bsube\label{eq10}
\be \label{tf1}
\hat f_{\alpha uk} = \hat f_{\alpha uk}^{+}+\hat f_{\alpha uk}^{-},
\ee
defined via
\be\label{van}
 \hat f_{\alpha u k}^{-}\rho^0_{\B}
=\rho^0_{\B}\hat f_{\alpha u k}^{+}=0.
\ee
\esube
This results in
\be
\begin{split}
&c_{\alpha uv k}^{-}(t)
\equiv \la\hat f_{\alpha u k}^{-;\B}(t)\hat f_{\alpha v k}^{+;\B}(0)\ra_{\B}=\eta^{-}_{\alpha u v k}e^{-\gamma_{\alpha k}t},
\\
&c_{\alpha uv k}^{+}(t) \equiv \la\hat f_{\alpha v k}^{-;\B}(0)\hat f_{\alpha u k}^{+;\B}(t)\ra_{\B}=\eta^{+}_{\alpha uv \bar k}e^{-\gamma_{\alpha k}t},
\end{split}
\ee
where $\eta^{-}_{\alpha u v k}\equiv \eta_{\alpha u v k}$
 and $\eta^{+}_{\alpha u v k}\equiv \eta_{\alpha u v \bar k}^{\ast}$.
As the thermofield excitation is concerned \cite{Ume95},
$\hat f_{\alpha u k}^{\pm}$ resembles
the creation/annihilation operator
onto the reference
$\rho^0_{\B}$
that participates in \Eqs{Fcorr_boson}
and (\ref{ff_boson}).

\noindent(\emph{iii})
 \emph{Thermofield Langevin ansatz}:
Each thermofield dissipaton satisfies
\be\label{QLE2}
\hat f_{\alpha u k}^{\pm}(t)=\hat f_{\alpha u k}^{\pm;\B}(t)
 \pm i\sum_{v}\!\int^{t}_{0}\!\d\tau
  c^{\pm}_{\alpha uvk}(t-\tau)\hat Q_v (\tau).
\ee
In compared with \Eq{QLE_boson}, the resolved are not only the absorptive versus emissive contributions, but also the Langevin force that reads $\hat f_{\alpha uk}^{\pm;\B}(t)=\hat f_{\alpha uk}^{\pm;\B}(0) e^{-\gamma_{\alpha k} t}$.
This recovers the generalized diffusion equation of the DEOM theory \cite{Yan14054105}, further including the thermal effects.

\subsection{System--bath entanglement theorem for correlation functions}
In the following, we elaborate above basic ingredients of the DTF theory, with a class of input--output relations between local and nonlocal nonequilibrium steady--state correlation functions.
Denote ${\bm C}_{\alpha\tS k}(t)=\{C_{\alpha \tS k}(t)\equiv \la \hat f_{\alpha u k}(t)\hat Q_{v}(0)\ra\}$ and ${\bm \phi}_{\alpha k}(t)=i[{\bm c}^{-}_{\alpha k}(t)-{\bm c}^{+}_{\alpha k}(t)]=\{\phi_{\alpha uvk}(t)= i[c_{\alpha uv k}^{-}(t)-c_{\alpha uv k}^{+}(t)]\}$. 
Equations (\ref{eq10})
and (\ref{QLE2}) give rise to
\be\label{ft_O_corr}
{\bm C}_{\alpha\tS k}(t)={\bm X}_{\alpha \tS k}(t)
-\int_{0}^t\!{\rm d}\tau\,{\bm \phi}_{\alpha k}(t-\tau)
 \cdot {\bm C}_{\tS\tS}(\tau)
\ee
where $X_{uvk}^{\alpha \tS}(t)=X_{uvk}^{\alpha \tS}(0)e^{-\gamma_{\alpha k}t}$, \Eq{g2}, and
\be\label{fO_ave}
{\bm X}_{\alpha \tS k}(0)
  \!=i\!\int_{0}^{\infty}\!\!{\rm d}\tau
   \big[{\bm c}_{\alpha k}^{+}(\tau) \cdot{\bm C}_{\tS\tS}^{\dg}(\tau)
    - {\bm c}_{\alpha k}^{-}(\tau) \cdot{\bm C}_{\tS\tS}^{T}(\tau)\big].
\ee
with ${\bm M}^{T}$ being  the matrix transpose.
Together with \Eq{ft_O_corr} and (\ref{g2}), we obtain further
\be\label{Xt_boson}
 {\bm X}_{\alpha\tS}(t)
=2\,{\rm Im} \! \int^{\infty}_0\!\!\d\tau\,
  {\bm c}_{\alpha}^{T}(t+\tau)\cdot{\bm C}_{\tS\tS}^{T}(\tau).
\ee
This completes  \Eq{CSalp_boson_final}, 
the system--bath entanglement theorem for nonequilibrium steady--state correlation functions.

The derivations of the key expression (\ref{fO_ave}) are as follows.
(i) Let us start with $\la \hat A(0)\ra
=\lim_{t\rightarrow\infty}{\rm Tr}\big[\hat A(t)\rho_{\T}^{\rm init}]$ for any operator $\hat A$.
This asymptotic identity holds
for any physically supported
initial total composite density
operator $\rho_{\T}^{\rm init}$.
In particular, we choose
$\rho_{\T}^{\rm init}=\rho_{\tS}^{\rm init}\otimes\rho^0_{\B}$, with $\rho^0_{\B}$ being the pure bath canonical ensemble density operator;
(ii) Then split $X_{uvk}^{\alpha \tS}(0)\equiv \la\hat f_{\alpha u k}(0)\hat Q_{v}(0)\ra=\la\hat f^{+}_{\alpha u k}(0)\hat Q_{v}(0)\ra+\la\hat Q_{v}(0)\hat f^{-}_{\alpha u k}(0)\ra$. This is true since the system and reservoir operators are  commutable  at any given local time;
(iii) Finally, obtain ${\rm Tr}[\hat f^{+}_{\alpha u k}(t)\hat Q_{v}(t)\rho_{\T}^{\rm init}]$
and ${\rm Tr}[\hat Q_{v}(t)\hat f^{-}_{\alpha u k}(t)\rho_{\T}^{\rm init}]$  from \Eq{QLE2},
with focus on their $t\rightarrow \infty$ expressions, where $\hat f^{\B;\pm}_{\alpha u k}(t)$ makes no contribution according to \Eq{van}.
The resulting
 $X_{uvk}^{\alpha \tS}(0)$ according to (ii)
is just \Eq{fO_ave}.

\subsection{Comments}
\begin{figure}
\includegraphics[width=1.0\columnwidth]{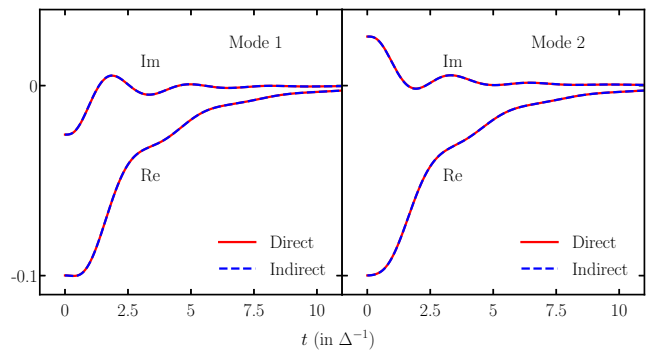}
\caption{
 Numerical validation of \Eq{ft_O_corr} with \Eq{fO_ave}  via the equality between lhs (direct) and rhs (indirect) of it,
exemplified with two different disspaton modes, (left and right panel). For simplicity, we adopt $ H_{\tS}=\frac{\Delta}{2}\hat \sigma_z+\frac{V}{2}\hat \sigma_x$ and $\hat Q=\Delta\hat \sigma_x$. The bath spectral density assumes $J(\w)=(\w_{\B}\zeta_{\B}\w)/[(\w_{\B}^2-\w^2)^2+\zeta_{\B}^2\w^2]$, which is related to the  interacting bath reservoir correlation functions in \Eq{Fcorr_boson} via the
fluctuation--dissipation theorem. Parameters are $V=\w_{\B}=\zeta_{\B}=k_BT=\Delta$.
}
\label{fig2}
\end{figure}

\paragraph{DTF versus DEOM.}
It is worth emphasizing that the present DTF formalism, \Eqs{eq9}--(\ref{QLE2}), is rather general in relation to the absorptive and emissive processes. Its application to obtain \Eqs{ft_O_corr}--(\ref{Xt_boson}) is an example that can be numerically verified by DEOM evaluations; see   \Fig{fig2} for the direct versus indirect calculations.  However, \Eqs{ft_O_corr}--(\ref{Xt_boson}) can not be obtained within the original DEOM framework.
%
That is to say, although both the DTF theory and DEOM method are numerically exact for Gaussian environments, DTF theory helps reveal more explicit relations.

Note that the present DTF goes with the exponential decomposition, \Eq{FFdecomp}; see in such as Ref.\,\onlinecite{Che22221102} for the latest developments of exponential decomposition methods.
Its generalization to the generic time--derivative closure decomposition scheme \cite{Xu07031107,Tan15224112,Hsi18014103,Hsi18014104,Cui19024110,Zha20064107,Ike20204101} 
can be readily constructed, similar to what we have done for the DEOM theory \cite{Gon20154111}.
%
%
Nevertheless, the exponential decomposition is often the choice towards the interpretation of experimental observations.
%
%
{\color{black}Besides, due to the obtained relation (\ref{ft_O_corr})--(\ref{Xt_boson}), we can compute ${\bm C}_{\alpha\tS k}(t)$ by only evaluating ${\bm C}_{\tS\tS}(t)$. The latter is much easier to converge in DEOM calculations. This will help save a lot of numerical costs.
}

\paragraph{DTF versus NEGF}
Furthermore, the $t=0$ behaviour of \Eq{CSalp_boson_final} with \Eq{Xt_boson}
is closely related to
the NEGF formalism of transport current \cite{Sch61407,Kel651018,Mei922512,Hau08,Gru1624514}.
For example, consider
the heat transport from the $\alpha$--reservoir
to the local impurity system.  The heat current operator reads
\be\label{hatI_boson_def}
 \hat J_{\alpha}
\equiv
 -\frac{{\rm d}h_{\alpha}}{{\rm d}t}=-i[H_{\T},h_{\alpha}]
 =\sum_{u}\dot{\hat F}_{\alpha u}\hat Q_{u}.
\ee
This is the electron transport analogue \cite{He18195437,Son17064308}.
The heat current is then [cf.\,\Eq{CalpS}]
\be\label{heat_current}
 J_{\alpha}=\sum_{u}\la\dot{\hat F}_{\alpha u}\hat Q_{u}\ra
=\sum_{u}\dot{C}^{\alpha\tS}_{uu}(t=0).
\ee
Now apply \Eq{CSalp_boson_final}, with noticing
that its second term does not
contribute to $\dot{C}^{\alpha\tS}_{uu}(0)$.
We obtain \cite{Du212155}
\be\label{heat_current_final}
 J_{\alpha}=2\,{\rm Im}\!\int^{\infty}_0\!\!\d \tau\,
  {\rm tr}\big [\dot{\bm c}_{\alpha}(\tau)\cdot{\bm C}_{\tS\tS}(\tau)\big].
\ee
This is the time--domain  Meir--Wingreen's formula  via the NEGF approaches \cite{Mei922512}. It is obtained in NEGF formalism by introducing the contour ordering,  followed by the use of Langreth’s rules.
%
Here, the DTF theory can produce it in a rather simple manner.
It is worth  noting that in the nonequilibrium scenaios, the definition of current in \Eq{hatI_boson_def} seems to be limited in the weak system--bath coupling \cite{Ren10170601}, but the polaron transformation can take into account of the system--bath interactions non-perturbatively beyond this limitation \cite{Xu16110308}.

\paragraph{DTF versus conventional thermofield approach}
Last but not least, we compare the DTF theory to the conventional thermofield approach \cite{Ume95}.
The thermofield theory has been extended to open quantum dynamics systems, and some studies have been conducted in combination with the hierarchical equations of motion formalism \cite{Ari91163,Ari91329,Kob03395,Bor19234102}. 
It goes by the purification of  the canonical thermal state, taking the single reservoir and single coupling mode case as an example,  as $|\xi \rangle=\prod_{j}\otimes|\xi_{ j} \rangle$, where 
$
|\xi_{  j} \rangle=(1/\sqrt{Z_{ j}})\sum_{n_{ j}}e^{-\beta_{ } n_{j} \omega_{ j}/2}|n_{ j}\rangle\otimes |n_{ j}
\rangle'$. With the partition function $Z_{ j}=(1-e^{-\beta \omega_{ j}})^{-1}$, the $|\xi_{ j} \rangle$ is a purification of the density operator
$
 \rho_{j}=e^{-\beta  \omega_{j} a_{ j}^{\dagger}a_{ j}}/Z_{j} = \sum_{n_{j}} (e^{-\beta  n_{  j}\omega_{ j}}/Z_{ j})|n_{ j}\rangle \langle n_{j}|$. Here $|n_{ j}\rangle \equiv \frac{1}{\sqrt{n_i !}}(a_{  j}^{\dagger})^{n_{ j}}|0_{  j}\rangle $ and $|n_{ j}\rangle' \equiv \frac{1}{\sqrt{n_{ j} !}}(b_{ j}^{\dagger})^{n_{ j}}|0_{ j}\rangle'$, with $a_{ j}^{\dagger}/a_{ j}$ and $b_{ j}^{\dagger}/b_{j}$ being the creation/annihilation operators of the original and assistant Fock space, respectively. The vacuum states are defined by
$ a_{ j}|0_{ j}\rangle=b_{ j}|0_{ j}\rangle'=0$.
It is easy to verify that $\rho_{ j}={\rm tr}'(|\xi_{  j} \ra\la\xi_{ j}|)$, where ${\rm tr}'$ represents the partial trace with respect to the assistant degrees of freedom.
Then, to obtain the zero temperature effective bath, we do the Bogoliubov transformation, which reads
\bsube
\begin{align} \label{bt1} a_{ j}=\sqrt{1+ \bar{n}_{  j}}c_{ j}+\sqrt{\bar{n}_{  j}}d_{  j}^{\dagger}, 
\\ \label{bt2} 
b_{  j}=\sqrt{1+ \bar{n}_{ j}}d_{  j}+ \sqrt{\bar{n}_{ j}}c_{j}^{\dagger},
\end{align}
\esube
where $\bar{n}_{ j}=(e^{\beta_{ }\omega_{  j}}-1)^{-1}$ is the average occupation number. 
It can be shown that 
$ c_{ j}|\xi_{ j} \rangle=d_{ j}|\xi_{  j}\ra =0, $
together with the bosonic commutation relation for $c_{j}^{\pm}$ and $d_j^{\pm}$.
Now we can add an assistant bath for each reservoir, which does not affect the original system--and--environment dynamics. It results in
$
  h'_{\B}=\sum_{j}\w_{ j} \big(a_{ j}^{\dg}a_{ j}-b_{ j}^{\dg}b_{  j}\big)=\sum_{j}\w_{ j} \big(c_{ j}^{\dg}c_{ j}-d_{ j}^{\dg}d_{  j}\big)
$.
After Bogoliubov transformation, \Eqs{bt1} and (\ref{bt2}), the total Hamiltonian becomes
$
H_{\T}=H_{\tS}+\hat Q \hat F+h'_{\B}
$, where
\be \label{ddd}
\hat F\equiv (1/\sqrt{2})\sum_j g_j(a_j+a_j^{\dg})=\hat F^{+}+\hat F^{-}, 
\ee
with 
\be\label{aiai} 
\hat F^{-}|\xi \rangle=\la \xi |\hat F^{+}=0.
\ee
Here, 
\be 
\hat F_{}^{\sigma}\equiv (1/\sqrt{2})\sum_j g_{  j}(\sqrt{1+ \bar{n}_{ j}}\hat c^{\sigma}_{ j}+\sqrt{\bar{n}_{ j}}\hat d^{\sigma}_{ j})
\ee
satisfies
\be \label{hhh}
\hat F_{}^{\sigma}(t)=\hat F_{}^{\sigma;\B}(t)-\sigma i\int_{0}^t\!\!{\rm d}\tau\,c^{\sigma}(t-\tau)\hat Q(\tau),
\ee
with $c^{+}(t)=[c(t)]^{\ast}=[c^{-}(t)]^{\ast}$ and
$
\hat F^{\sigma;\B}(t)\!=\sum_j(g_{ j}/\sqrt{2}) (\!\sqrt{1+ \bar{n}_{j}}\hat c^{\sigma}_{j}e^{\sigma i\w_{j}t}\!+\!\!\sqrt{\bar{n}_{ j}}\hat d^{\sigma}_{j}e^{-\sigma i\w_{j}t})
$.
 Comparing between the conventional thermofield and the DTF formalism,  we may observe
 \Eqs{eq9}--(\ref{eq10}) with \Eq{QLE2} as the statistical quasi-particle analogue to \Eqs{ddd}--(\ref{aiai}) with \Eq{hhh}.
 The present theory would be better physically supported since the 
introduced discrete statistical quasi-particle picture, with the Langevin force in \Eq{QLE2} being effectively resolved.
 This agrees with the generalized diffusion equation of the DEOM theory \cite{Yan14054105}.

\section{Concluding remarks}
In conclusion, we develop
a statistical quasi-particle thermofield theory, the DTF theory, which is exact 
assuming the Gaussian influence environments.
It goes with the dissipaton decomposition of the hybrid bath reservoir mode.
The DTF theory bridges the NEGF and the real--time reduced system dynamics methods.
Universal relations for
a class of important nonequilibrium steady--state correlation functions are established.
The fermionic counterparts to
the present DTF theory and the resulting nonequilibrium system--bath entanglement theorem
can be readily established in a similar manner.
It is also interesting to investigate its deep relationship to the stochastic formalism of quantum dissipation, where the quantum noise appears naturally and the moments of noise lead to a set of generalized hierarchical equations \cite{Hsi18014103,Hsi18014104}.
Generally speaking, the DTF theory is an important ingredient in the study of open
quantum systems and will serve
as a versatile tool to such as the nonequilibrium thermodynamics
and  transport phenomena  in strongly correlated systems.

%
%
%


\begin{acknowledgments}
Support from
the Ministry of Science and Technology of China (Grant No.\ 2017YFA0204904 and 2021YFA1200103) and the National Natural Science Foundation of China (Grant Nos.\ 22103073 and 22173088) and the Anhui Initiative in Quantum Information Technologies is gratefully acknowledged. Y. Wang and Z. H. Chen thank also the
partial support from GHfund B (20210702).
We are indebted to Peng-Li Du, Xiang Li, and  Yu Su for valuable discussions.
 \end{acknowledgments}

The data that support the findings of this study are available from the corresponding author upon reasonable request. 


\end{document}